# Underground operation of the ICARUS T600 LAr-TPC: first results

C. Rubbia[a,j], M. Antonello[a], P. Aprili[a], B. Baibussinov[b], M. Baldo Ceolin[b], L. Barzè[c], P. Benetti[c], E. Calligarich[c], N. Canci[a], F. Carbonara[d], F. Cavanna[e], S. Centro[b], A. Cesana[f], K. Cieslik[g], D. B. Cline[h], A. G. Cocco[d], A. Dabrowska[g], D. Dequal[b], A. Dermenev[i], R. Dolfini[c], C. Farnese[b], A. Fava[b], A. Ferrari[j], G. Fiorillo[d], D. Gibin[b], A. Gigli Berzolari[c], S. Gninenko[i], T. Golan[k], A. Guglielmi[b], M. Haranczyk[g], J. Holeczek[l], P. Karbowniczek[g], M. Kirsanov[i], J. Kisiel[l], I. Kochanek[l], J. Lagoda[m], M. Lantz[f], S. Mania[l], G. Mannocchi[n], F. Mauri[c,f], A. Menegolli[c], G. Meng[b], C. Montanari[c], S. Muraro[f], S. Otwinowski[h], O. Palamara[a], T. J. Palczewski[m], L. Periale[n], A. Piazzoli[c], P. Picchi[n], F. Pietropaolo[b,*], P. Plonski[o], M. Prata[c], P. Przewlocki[m], A. Rappoldi[c], G. L. Raselli[c], M. Rossella[c], P. Sala[f], E. Scantamburlo[e], A. Scaramelli[f], E. Segreto[a], F. Sergiampietri[p], J. Sobczyk[k], D. Stefan[a,g], J. Stepaniak[m], R. Sulej[m], M. Szarska[g], M. Terrani[f], F. Varanini[b], S. Ventura[b], C. Vignoli[a], T. Wachala[g], H. Wang[h], X. Yang[h], A. Zalewska[g], K. Zaremba[o], J. Zmuda[k]

[a] *Laboratori Nazionali del Gran Sasso dell'INFN, Assergi (AQ), Italy*
[b] *Dipartimento di Fisica e INFN, Università di Padova, Via Marzolo 8, I-35131, Padova, Italy*
[c] *Dipartimento di Fisica Nucleare e Teorica e INFN, Università di Pavia, Via Bassi 6, I-27100, Pavia, Italy*
[d] *Dipartimento di Scienze Fisiche, INFN e Università Federico II, Napoli, Italy*
[e] *Dipartimento di Fisica, Università di L'Aquila, via Vetoio Località Coppito, I-67100, L'Aquila, Italy*
[f] *INFN, Sezione di Milano e Politecnico, Via Celoria 2, I-20123, Milano, Italy*
[g] *The Henryk Niewodniczanski, Institute of Nuclear Physics, Polish Academy of Science, Krakow, Poland*
[h] *Department of Physics and Astronomy, University of California, Los Angeles, USA*
[i] *INR RAS, prospekt 60-letiya Oktyabrya 7a, Moscow 117312, Russia*
[j] *CERN, Ch1211 Geneve 23, Switzerland*
[k] *Institute of Theoretical Physics, Wroclaw University, Wroclaw, Poland*
[l] *Institute of Physics, University of Silesia, 4 Uniwersytecka st., 40-007 Katowice, Poland*
[m] *A. Soltan Institute for Nuclear Studies, 05-400 Swierk/Otwock, Warszawa, Poland*
[n] *Laboratori Nazionali di Frascati (INFN), Via Fermi 40, I-00044, Italy*
[o] *Institute for Radioelectronics, Warsaw Univ. of Technology Pl. Politechniki 1, 00-661 Warsaw, Poland*
[p] *Dipartimento di Fisica, Università di Pisa, Largo Bruno Pontecorvo 3, I-56127, Pisa, Italy*

ABSTRACT: Open questions are still present in fundamental Physics and Cosmology, like the nature of Dark Matter, the matter-antimatter asymmetry and the validity of the particle interaction Standard Model. Addressing these questions requires a new generation of massive particle detectors exploring the subatomic and astrophysical worlds. ICARUS T600 is the first large mass (760 ton) example of a novel detector generation able to combine the imaging capabilities of the old famous "bubble chamber" with an excellent energy measurement in huge electronic detectors. ICARUS T600 now operates at the Gran Sasso underground laboratory, studying cosmic rays, neutrino oscillation and proton decay. Physical potentialities of this novel telescope are presented through few examples of neutrino interactions reconstructed with unprecedented details. Detector design and early operation are also reported.

---

[*] Corresponding author. *E-mail*: `francesco.pietropaolo@cern.ch`

# Contents

# 1. Introduction

Historically, imaging detectors have always played a crucial role in particle physics. In the past century successive generations of detectors realized new ways to visualize particle interactions driving the advance of the physical knowledge and the discovery of unpredicted phenomena, even on the basis of single fully reconstructed events. In particular, bubble chamber detectors were an incredibly fruitful tool permitting to visualize and measure particle interactions, providing fundamental contributions to particle physics discoveries. Gigantic bubble chambers like Gargamelle [1] (3 ton of mass) were extraordinary achievements, successfully employed in particular in neutrino physics. The major limitations of bubble chambers in the search for rare phenomena are the impossibility to scale their size towards much larger masses and their duty cycle intrinsically limited by the mechanics of the expansion system.

In 1977 C. Rubbia [2] conceived the idea of a Liquid Argon Time Projection Chamber (LAr-TPC), i.e. calorimetric measurement of particle energy together with three-dimensional track reconstruction from the electrons drifting in an electric field in sufficiently pure Liquid Argon. The LAr-TPC successfully reproduces the extraordinary imaging features of the bubble chamber - its medium and spatial resolution being similar to those of heavy liquid bubble chambers - with the further achievement of being a fully electronic detector, potentially scalable to huge masses (~ many kt). In addition the LAr-TPC provides excellent calorimetric measurements and has the big advantage of being continuously sensitive and self-triggering.

The ICARUS T600 cryogenic detector is the biggest LAr-TPC realized ever, with the cryostat containing 760 tons of LAr. Its construction finalized many years of ICARUS Collaboration R&D studies, with bigger and bigger in mass laboratory and industrial prototypes. Nowadays, it represents the status of the art of this technique and it marks a major milestone in the practical realization of large-scale LAr detectors. The ICARUS T600 is now installed in the Hall B of the Gran Sasso underground National Laboratory (LNGS) of Istituto Nazionale di Fisica Nucleare (INFN), shielded against cosmic rays by about 1400 meters of rock. Smoothly running under stable conditions since months, the detector is demonstrating high-level technical performances, as will be shown in the following.

The ICARUS T600 addresses a wide physics program operating as a continuously sensitive general-purpose observation instrument. It is simultaneously collecting a wide variety of "self-triggered" events of different nature, namely cosmic ray events (atmospheric and solar neutrino interactions), charge and neutral current neutrino interactions associated with the CNGS neutrino beam, focusing on neutrino oscillation search and in particular on $\nu_\mu \to \nu_\tau$ appearance. It will also be searching for rare events up to now unobserved like the long sought for proton decay with zero background in one of its $3\times10^{32}$ nucleons (in particular into exotic channels). Relying on the LAr-TPC three-dimensional, high granularity imaging and calorimetric capabilities, few proton decay events may be enough to discover matter instability. ICARUS T600 could also provide interesting results on supernova explosion mechanism in case of a galactic supernova event.

The ICARUS T600 detector description and its underground operation in 2010 in the LNGS are presented in this paper. A couple of fully reconstructed neutrino interactions, collected by the ICARUS T600 during its initial operation underground, will be discussed, demonstrating the level of detail and the power this detection technique.

## 2. The ICARUS T600 detector

The ICARUS T600 detector consists of a large cryostat split into two identical, adjacent half-modules with internal dimensions $3.6 \times 3.9 \times 19.6$ m$^3$ and filled with about 760 tons of ultra-pure liquid Argon [3]. A uniform electric field ($E_D$ = 500 V/cm) is applied to the LAr bulk: each half-module houses two TPCs separated by a common cathode.

Charged particles, generated for example by a neutrino interaction in LAr, crossing the medium produce ionization along their path. Thanks to the low transverse diffusion of charge in LAr, the electron images of ionization tracks are preserved and, drifting along the electric field lines, are projected onto the anode (TPC) – see the illustration in figure 1.

Each TPC is made of three parallel planes of wires, 3 mm apart, facing the drift path (1.5 m). Globally, 53248 wires with length up to 9 m are installed in the detector. By appropriate voltage biasing, the first two planes (Induction-1 and Induction-2 planes) provide signals in non-destructive way, whereas the ionization charge is finally collected by the last one (Collection plane).

Wires are oriented on each plane at a different angle (0°, +60°, -60°) with respect to the horizontal direction. Therefore, combining the wire coordinate on each plane at a given drift time, a three-dimensional image of the ionizing event can be reconstructed. A remarkable resolution of about 1 mm$^3$ is uniformly achieved over the whole detector active volume (~170 m$^3$).

The measurement of the absolute time of the ionizing event, combined with the electron drift velocity information ($v_D \sim 1.6$ mm/μs at $E_D$ = 500 V/cm), provides the absolute position of the track along the drift coordinate. The determination of the absolute time of the ionizing event is accomplished by prompt detection of the scintillation light produced in LAr by charged particles. For this purpose arrays of Photo Multiplier Tubes (PMTs), suitable to detect VUV scintillation light ($\lambda$ = 128 nm) and operating at the LAr cryogenic temperature [4], are installed behind the wire planes.

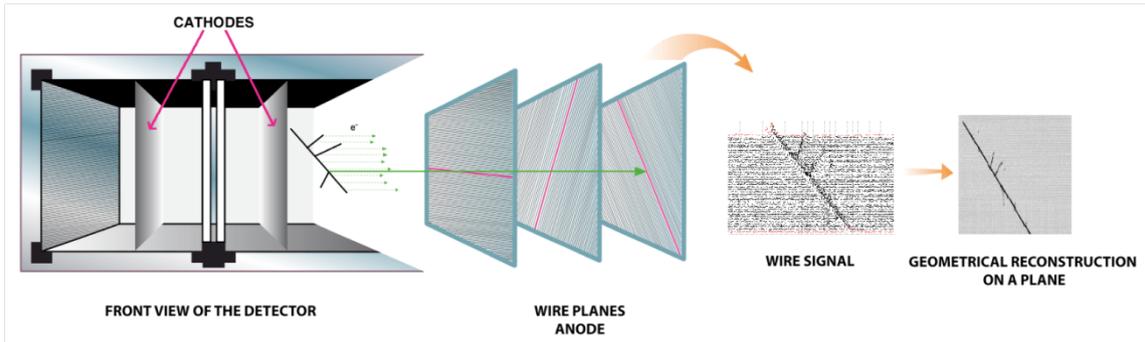

**Figure 1. Illustration of the ICARUS T600 working principle: from a charged particle ionization path in LAr to its geometrical reconstruction.**

The electronics was designed to allow continuous read-out, digitization and independent waveform recording of signals from each wire of the TPC. The read-out chain is organized on a 32-channel modularity. A Decoupling Board receives the signals from the chamber and passes them to an Analogue Board via decoupling capacitors; it also provides wire biasing voltage and the distribution of the test signals. The Analogue Board hosts the front-end amplifiers and performs 16:1 channel multiplexing and 10-bit ADC digitization at 400 ns sampling time per channel. The overall gain is about 1000 electrons per ADC count, setting the signal of minimum ionizing particles to ~ 15 ADC counts with a dynamic range of about 100 minimum ionizing particles. A 3 µs decay constant is used for the unipolar signals coming Collection and Induction 1 wires, while 100 µs decay constant is used for the bipolar current signals (Induction 2 wires). A Digital Board hosts as a 10 bit wide, waveform recorder. It continuously reads the data, stores them in multi-event circular buffers, each covering a full drift distance. When a trigger signal occurs, the active buffer is frozen, writing operations are moved to the next free buffer, and the stored data are read out by the DAQ. This configuration guarantees no dead time, until the maximum DAQ throughput (1 full-drift event per second) is reached. The average electronic noise achieved with the special designed low noise front-end is well within expectations: 1500 electrons r.m.s. to be compared with ~ 15000 free electrons produced by a minimum ionizing particle in 3 mm (S/N ~ 10).

The ICARUS T600 detector was pre-assembled since 1999 in Pavia (Italy), where one of its two 300-tons half-modules was brought to operation in 2001. A test run of three months was carried out with cosmic rays at the Earth surface, allowing for the first time an extensive study of the main detector features [5]. After the test, the detector was de-commissioned and, in 2004, the two cryostats housing the inner detector were transported to their final site, in the Hall B of the underground Gran Sasso National Laboratories.

A number of activities on the ICARUS T600 plant was then necessary for the assembling of the detector in its underground site, and in the first months of 2010 the T600 module was finally brought into operation with its commissioning and the detection of the first events from the CNGS neutrino beam and cosmic rays.

In order to fulfill safety and reliability requirements for underground operation at LNGS, the ICARUS T600 module, sketched in figure 2, was equipped with dedicated technical infrastructures.

One thermal insulation vessel surrounds the two half-modules: it is realized with evacuated honeycomb panels assembled to realize a tight containment. Between the insulation and the aluminum containers a thermal shield is placed, with boiling Nitrogen circulating inside to intercept the heat load and maintain the cryostat bulk temperature uniform (within 1 K) and stable at 89 K.

Nitrogen used to cool the T300 half-modules is stored into two 30 m$^3$ LN2 tanks. Its temperature is fixed by the equilibrium pressure in the tanks (~ 2.1 bar, corresponding to about 84 K), which is kept stable in steady state by a dedicated re-liquefaction system of twelve cryo-coolers (48 kW global cold power), thus guaranteeing a safe operation in closed-loop.

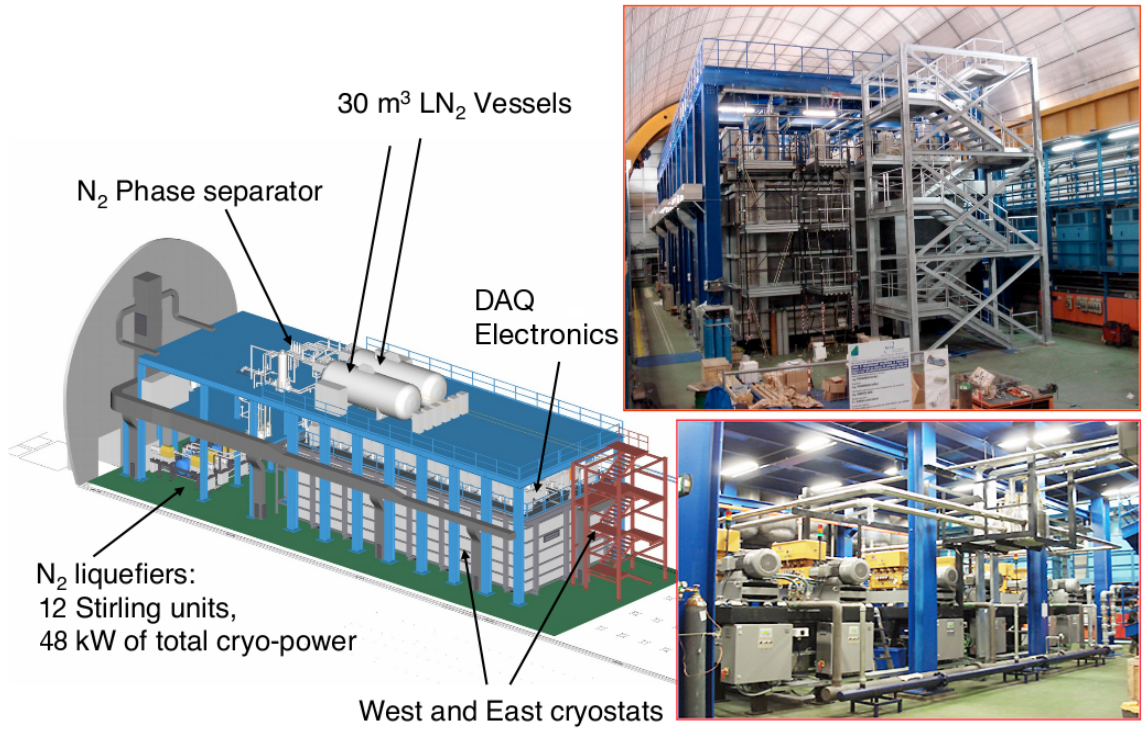

**Figure 2. Left: Schematic view of the whole ICARUS T600 plant in HALL-B at LNGS. Right-top: photo of the actual detector installation. Right-bottom: details of the cryo-cooler plant.**

A fundamental requirement for the performance of a LAr-TPC is that electrons produced by ionizing particles can travel "unperturbed" from the point of production to the wire planes. In other words, electronegative impurities (mainly $O_2$, $H_2O$ and $CO_2$) in LAr must be kept at a very low concentration level (less than 0.1 ppb). To this aim, each half-module is equipped with two gas argon and one liquid argon recirculation/purification systems. Argon gas is continuously drawn from the cryostat ceiling and, re-condensed, drops into Oxysorb ™ filters to finally get back into the LAr containers. LAr instead is recirculated by mean of an immersed, cryogenic pump (~ 2 m$^3$/h, full volume recirculation in 6 days) and is purified through standard Hydrosorb/Oxysorb™ filters before being re-injected into the cryostats.

To ensure an acceptable initial LAr purity, before filling, the cryostats were evacuated to a pressure lower than 10$^{-4}$ mbar. Vacuum phase lasted for three months. In 7 days cryostats were then cooled down to a temperature of 90 K. Finally they were filled, in about two weeks, with commercial LAr, purified in-situ before entering the detector, at a rate of ~ 1 m$^3$/hour/cryostat. During the whole period the four gaseous re-circulations were operating at maximum speed to intercept the degassing impurities. One month after filling, the forced liquid argon recirculation and purification started on both cryostats at a rate of ~ 1 m$^3$/hour/cryostat.

## 3. Operations of ICARUS-T600 at LNGS

During summer 2010 the first events from the CNGS neutrino beam and cosmic rays were triggered. The trigger system relies on both, the scintillation light signals provided by the internal PMTs and by the CNGS proton extraction time. As a starting layout, for each of the four chambers, the analog sum of signal from PMTs is exploited with a discrimination threshold set at around 100 photo-electrons, guaranteeing an almost full efficiency for the interactions induced by CNGS neutrinos. The trigger for the CNGS neutrino interactions is based on the presence of the PMT signal within a CNGS related gate. For every CNGS SPS cycle two proton spills, lasting 10.5 µs each, separated by 50 ms, are extracted from the SPS machine. An "early warning" signal is sent from CERN to LNGS 80 ms before the first proton spill extraction, allowing the opening of two ~ 60 µs gates in correspondence to the predicted extraction times. The most accurate timing inside the controller is realized by a 40 MHz counter, reset every millisecond by a synchronization signal containing absolute time information, generated by the master clock unit of the LNGS external Laboratories (GPS unit and 10 MHz atomic clock) synchronized to CERN SPS accelerator clock. A trigger rate of about 1 mHz is obtained including neutrino interactions inside the detector and muons from neutrino interactions in the upstream rocks. The analysis of the recorded neutrino interactions in LAr, shows a synchronization between ICARUS and the actual proton extraction time as written in the CNGS-database, sufficient to reconstruct the 10.5 µs width of the two proton spills (figure 3). The residual 2.4 ms delay is in agreement with the neutrino time of flight (2.44 ms) taking into account the timing signal propagation delay to Hall B (~ 44 µs).

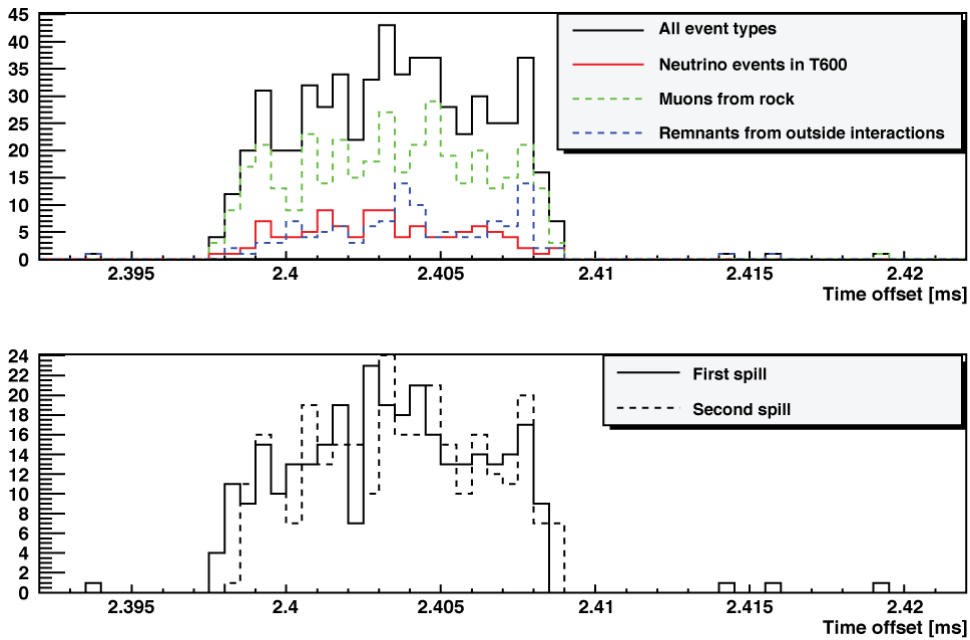

**Figure 3. Time distributions of the recorded neutrino interactions in the T600 with respect to the CNGS spill time (for 4.2 10$^{18}$ pot). Top: according to event classification; bottom: separately for first and second CNGS spill.**

The trigger for cosmic rays exploits the coincidence of the PMT's sum signals of the two adjacent chambers in the same half-module, relying on the 50% transparency of the cathode mechanical structure. This allows for an efficient reduction of the spurious signals maximizing the detection of low energy events. A trigger rate of 25 mHz has been achieved well below the maximum allowed DAQ rate, resulting in about 83 cosmic events per hour collected in the full T600.

As previously mentioned, liquid Argon purity and its measurement are key issues for the detector imaging capability and for a correct estimation of the energy deposition from the ionization charge signal. Purity is continuously monitored measuring the charge attenuation along ionizing cosmic muon tracks crossing the full drift path without evident associated δ-rays and γ's. About 50 muon tracks are sufficient to measure day-by-day the electron charge attenuation within 3% precision dominated by residual Landau charge fluctuations.

With the liquid recirculation turned on, the LAr purity steadily increased, reaching values of free electron lifetime[1] ($\tau_{ele}$) exceeding 6 ms in both half-modules after few months of operation (figure 4). In the East half-module this value was reached after tuning of the recirculation circuit. This corresponds to a maximum free electron yield attenuation of 16%, at the maximum drift distance of 1.5 m (corresponding to 1 ms maximum drift time). Pump maintenance required some stops of the LAr recirculation lasting several days, resulting in a sudden degradation of the purity, however the $\tau_{ele}$ went never below 1 ms. The different asymptotic levels of LAr purity, reached after each restarting of the recirculation system, are probably due to different equilibrium values of the effective recirculation speed.

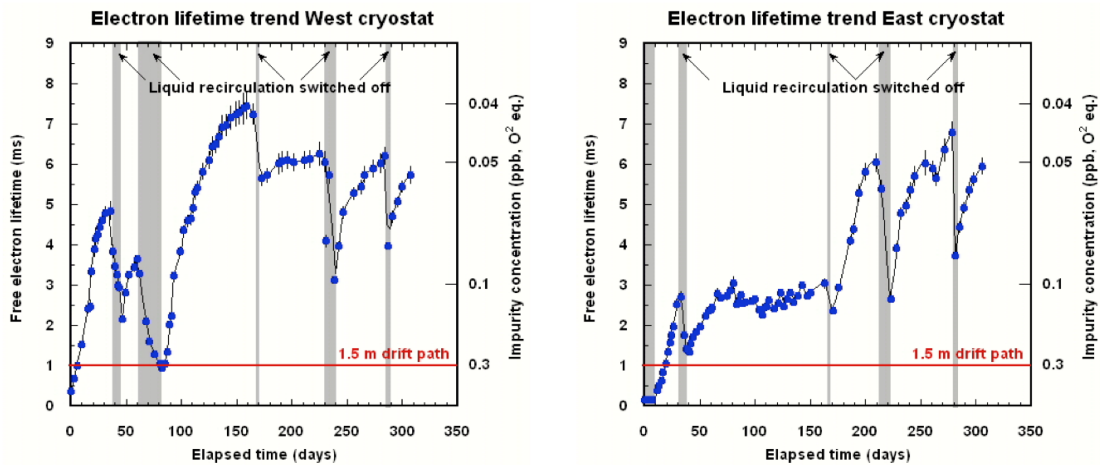

**Figure 4. Free electron lifetime evolution in the West (left) and East (right) cryostats as a function of the elapsed time for the first 350 days of operation of the T600 detector. For details see the text.**

The evolution of the residual impurity concentration can be described with a simple model as $dN(t)/dt = -N/\tau_R + k_L + k_D \exp(-t/\tau_D)$ where $\tau_R$ is the time needed to recirculate a full detector volume, $k_L$ is the total impurity leak rate and degassing rate and $k_D$ is the internal residual degassing rate assumed to vanish with a time constant $\tau_D$. Uniform distribution of the impurities throughout the detector volume is also assumed, as experimentally supported by the lifetime measurement with muon tracks in different regions of the TPC's.

---

[1] Free electron lifetime is the average capture time of a free ionization electron by an electronegative impurity in LAr.

Expressing the free electron lifetime as $\tau_{ele}$ *[ms] ~ 0.3/N [ppb]*, where the concept of Oxygen equivalent impurities is used, and fitting the data with this model yields a recirculation time of 6 days, in agreement with the nominal pump speed, and extremely low leak rates (< few ppt/day $O_2$ equivalent impurity concentration) in both half-modules. Internal initial degassing rate seems to be higher in the East module, resulting in a slightly worse purification rate.

## 4. Observation and reconstruction of neutrino events

The LAr-TPC performances have been studied progressively over the last two decades exposing different detectors to cosmic rays and neutrino beams, culminating with the successful achievement of the T600 operation [6] [7] [8] [9]. The high resolution and granularity of the detector imaging allow for a precise reconstruction of events topology and the recognition of the particles produced in interactions in LAr. The event reconstruction is completed by calorimetric measurement via dE/dx ionization signal over a very wide energy range, from MeV to several tens of GeV.

Identification of the nature of particles is obtained by studying the event topology and the energy deposition per track length unit as a function of the particle range (dE/dx versus range) for muons/pions, kaons and protons. A dedicated reconstruction program based on the polygonal line algorithm [10] for 3D reconstruction, and on neural network for particle identification was used to this purpose. Electrons are fully identified by the characteristic electromagnetic showering, well separated from $\pi^0$ via $\gamma$ reconstruction, dE/dx signal comparison and $\pi^0$ invariant mass measurement at the level of $10^{-3}$. This feature guarantees a powerful identification of the CC electron neutrino interactions, while rejecting NC interactions to a negligible level.

The electromagnetic energy resolution *$\sigma(E)/E = 0.03/\sqrt{(E(GeV))} \oplus 0.01$* is estimated in agreement with the $\pi^0 \to \gamma\gamma$ invariant mass measurements in the sub-GeV energy range [11]. The measurement of the Michel electron spectrum from muon decays, where bremsstrahlung photons emission is taken into account [12], provided the energy resolution below critical energy ($E_c \sim 30$ MeV), *$\sigma(E)/E = 0.11/\sqrt{(E(MeV))} \oplus 0.02$*. At higher energies the estimated resolution for hadronic showers is *$\sigma(E)/E = 0.30/\sqrt{(E(GeV))}$*. However the LAr-TPC detector allows to identify and measure, track by track, each hadron produced in interactions, through ionization and range, leading to a much better energy resolution.

For long muon track escaping the detector, momentum is determined exploiting the multiple scattering along the track, studying its displacements with respect to a straight line. The procedure, implemented as a Kalman filter technique and validated on cosmic rays stopping muons, allows a resolution $\Delta p/p$ that can be as good as 10%, depending mainly on the track length [13].

Examples of both CNGS beam and atmospheric neutrino interactions detected by the ICARUS T600 are presented below.

A CNGS $\nu_\mu$ CC event with a 13 meters long muon track is shown in figure 5 together with zoomed projections for Collection and Induction-2 views. The deposited energy associated to the single visible tracks corrected for quenching [14] and electron attenuation along the drift are collected in table 1 together with their directions in space. The use of two different views allows for recognizing the presence of two distinct electromagnetic showers pointing to the primary vertex. The corresponding conversion distances are measured to be 23 mm and 69 mm respectively. Even if the two photons overlap in the Collection view, used for the calorimetric energy measurement, the associated invariant mass $m_{12}^* = 125\pm15$ MeV/c$^2$, compatible with the $\pi^0$ mass, is determined, under the assumption of equally shared energy between the two photons, from their opening angle $\theta_{12} = (27.7\pm2.5)°$ and the total energy $E_{tot} = 521\pm16$ MeV. The closer photon initial ionization – averaging over the first 6 hits – amounts to 2.2 m.i.p. (minimum ionizing particle): this is a clear signature of a pair conversion confirming the expected e/$\gamma$ identification capabilities of the detector.

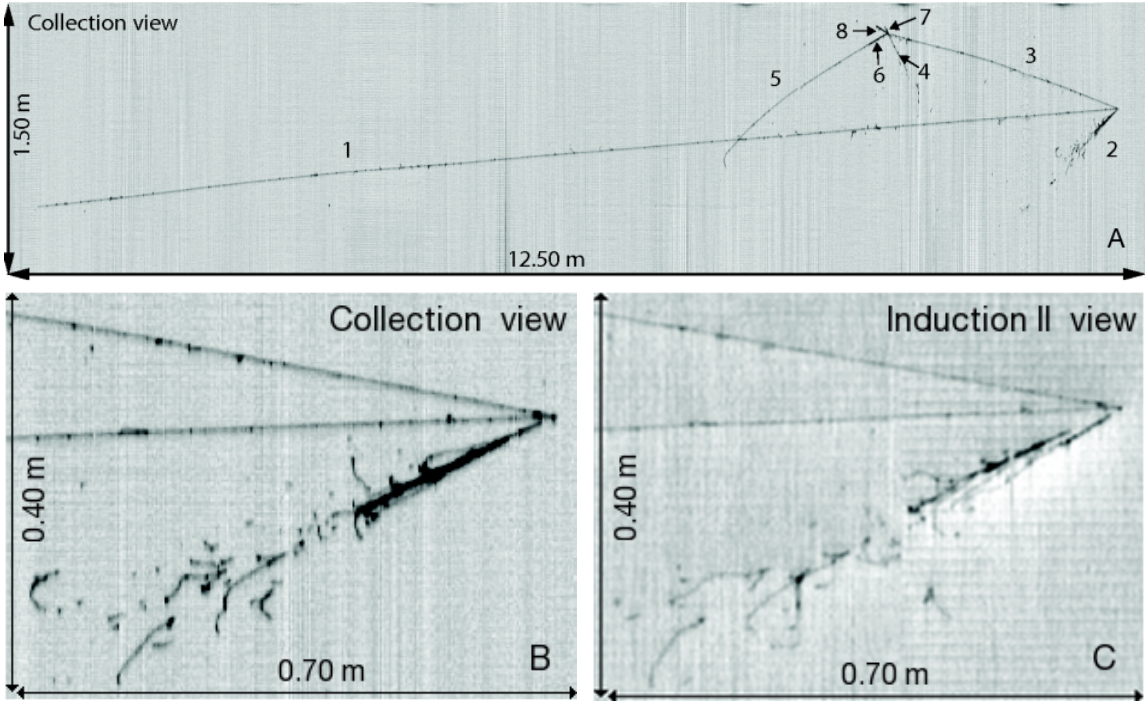

**Figure 5. An example of $\nu_\mu$ CC interaction from the CNGS beam in Collection view [A]; a close-up view of the primary vertex for the Collection [B] and Induction-2 [C] projections allow solving completely the event topology.**

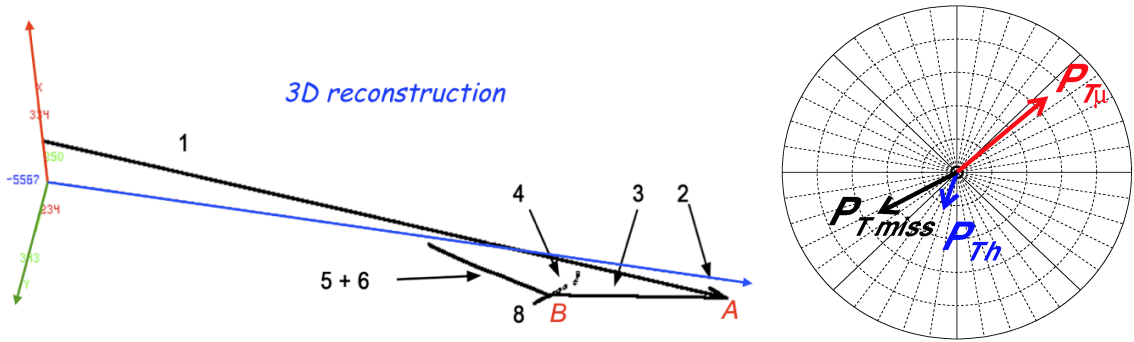

**Figure 6. Left: 3D reconstruction of the event shown in figure 5. Right: Momentum reconstruction in the transverse plane for muon ($P_{T\mu}$), hadrons ($P_{Th}$) and missing ($P_{Tmiss}$);**

– 8 –

the neutrino direction points into the page; the outermost circle corresponds to 500 MeV/c.

The leading muon deposits by ionization a total energy of about 2.7 GeV before exiting the detector. Due to multiple scattering, the muon track deviates up to 66 mm from a straight line through its first and the last points. Thanks to the 200 μm resolution measured along the drift coordinate on each wire, exploiting multiple scattering the momentum of the long muon track has been measured to be $p_\mu$ = 10.5±1.1 GeV/c.

The other primary long track is identified as a pion that interacts giving a secondary vertex. Its kinetic energy is assumed to be equal to the total energy deposited by track 3 and its daughters. Track 5 is identified as a kaon, decaying in flight into a muon. The kaon momentum at the decay point has been calculated from the muon energy and the decay angle. Adding the measured energy deposition before decay, the initial kaon momentum can be evaluated as 672±44 MeV/c. The capability to identify and reconstruct low energy kaons is a major advantage of the LAr-TPC technique for proton decay searches. In this event, the kaon momentum is not far from the average one (300 MeV/c) for instance, in the $p \to \nu K^+$ channel. Also the ability to identify $\pi^0$, as in this event, is important for many nucleon decay channels, as well as for the discrimination of neutral current events when looking for $\nu_\mu \to \nu_e$ oscillations.

**Table 1. Parameters of tracks visible in figure 5. The direction cosines are relative to a Cartesian system having the *z* axis parallel to the CNGS beam, the *y* axis parallel to the drift direction and the *x* axis pointing upwards.**

| Track | $E_{dep}$ (MeV) | cosx | cosy | cosz | PID |
|---|---|---|---|---|---|
| 1 | 2702 | 0.021 | 0.024 | 0.999 | μ, P=10.5 GeV/c |
| 2 | 521 ±16 | 0.002 | 0.420 | 0.908 | $\pi^0$ |
| 3 | 753 ±64 | -0.057 | -0.137 | 0.989 | π |
|   | 876 | From secondary vertex | | | |
| 4 | 119 ±10 | 0.052 | 0.649 | -0.759 | Multiple tracks |
| 5 | 436 ±37 | -0.086 | 0.251 | 0.964 | μ+e, from 6 |
| 6 | 184 ±16 | -0.055 | 0.239 | 0.969 | K |
| 7 | 54 ±5 | 0.388 | -0.793 | 0.469 | proton |
| 8 | 83 ±7 | -0.656 | -0.150 | 0.740 | escaping |

The total hadronic energy in the event, $E_h$ = 2.3±0.5 GeV, is certainly underestimated, since one particle (marked as 8 in figure 5-A) from the secondary vertex escapes from the bottom face of the detector, and neutrons may have been produced and escaped detection. The missing transverse momentum reconstructed for the event is 250 MeV/c; it is represented in figure 6 together with the 3D reconstruction of the event. Despite the non-full containment of the event, the missing transverse momentum value is fully consistent with the theoretical expectation from the Fermi motion of target nucleons, and compatible with the measurements at the WANF neutrino beam with the 50-liter LAr-TPC prototype [15]. The reconstructed total energy is 12.6±1.2 GeV, within the energy range of the CNGS beam [16].

A 6-prong atmospheric neutrino interaction confined in less than 1 m$^3$ of liquid Argon is shown in figure 7 (2D views) and figure 8 (3D reconstruction). It was recorded several seconds off the CNGS proton spill.

Table 2 gives the particle identification and the reconstructed energy of the tracks according to labels in figure 7. The total deposited energy results to be 910 MeV while the total reconstructed momentum is 800±90 MeV/c, in a direction at about 35 degrees from the CNGS beam direction. Several isolated hits and small clusters are visible, corresponding to energy depositions of few MeV each. They can come from three sources: bremsstrahlung photons, nuclear de-excitation γ from the primary vertex, and γ produced by neutron inelastic scattering, by neutron(s) generated in the primary vertex. Track 3 is identified as a muon from the interaction vertex by dE/dx versus range, hence the event is identified as a $\nu_\mu$ CC interaction.

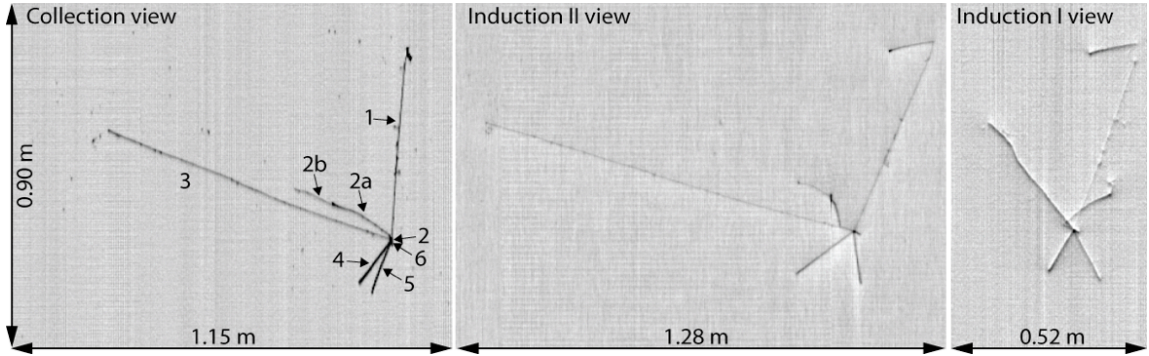

**Figure 7.** Example of low energy neutrino interaction. Three different 2D views are shown.

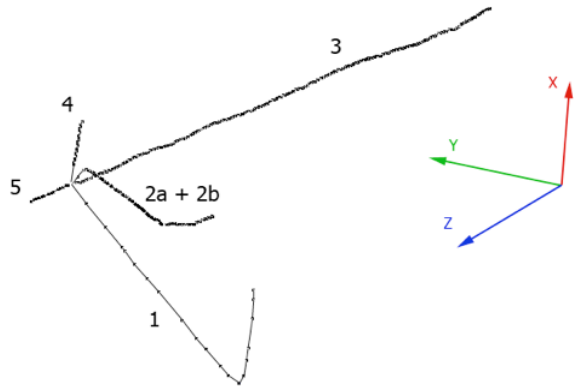

**Figure 8.** Three-dimensional reconstruction of the event shown in figure 7.

**Table 2.** Parameters of tracks visible in figure 7. Track 1a is the small one at the end of track 1.

| Track | $E_k$ (MeV) | Range (cm) | PID | |
|---|---|---|---|---|
| 1 | 130[1] ±11 | 49.3 | π | |
| 1a | 102 ±9 | 12.6 | π | |
| 2 | 26[1] | 3.3 | | decays into 2a |
| 2a | 125 ±11 | 19.9 | μ | decays at rest into 2b |
| 2b | 21 ±2 | 10.2 | e | |
| 3 | 227 ±19 | 104.9 | μ | |
| 4 | 101 ±9 | 13.5 | p | merged with 5 |
| 5 | 138±12 | 14.2 | p | |
| 6 | | 2.9 | – | merged with others |
| [1] Deposited energy along the track only | | | | |

## 5. Conclusions

The ICARUS T600 detector is so far the biggest LAr detector ever built. It has been successfully installed in the Gran Sasso underground laboratory and it is presently collecting data after having smoothly reached the optimal working conditions. LAr is a cheap liquid vastly produced by industry, which potentially permits to realize large mass detectors. ICARUS T600 represents the final milestone of a series of fundamental technological achievements in the last several years; its underground operation demonstrates that the ICARUS technology is now mature and scalable to much larger masses, in the range of tens of kton [17] as required to realize the next generation experiments for neutrino physics and proton decay searches. Finally, the examples of neutrino interaction event analyzed in this paper demonstrate that also the reconstruction procedure is well under control fully exploiting the physical potentiality of this technology.

## Acknowledgments

The ICARUS Project T600 has passed, in more than twenty years, through different stages of R&D and different prototypes. The final successful result is due to the continuous support of institutions and people who devoted their efforts to various aspects of this challenging project. ICARUS T600 could not be achieved without the fundamental financial contribution of INFN and encouragement of its presidents that have followed over the years: Luciano Maiani, Enzo Iarocci and Roberto Petronzio. The approval and encouragement of the Scientific Committee INFN, in the persons of its presidents Furio Bobisut, Carlo Bemporad, Francesco Ronga and Roberto Battiston have been essential. The LNGS with its directors Alessandro Bettini, Eugenio Coccia and Lucia Votano, along with all the technical staff of the Laboratories, gave a fundamental contribution. The Russian group warmly acknowledges the support of the INR director Victor Matveev and his involvement in the experiment. During this period many valuable colleagues have made different choices and have left the experiment deciding to follow different routes after having left relevant scientific and technical contributions. We want to remind the researchers of IHEP of Beijing, led by Ma-Ji Mao, those of ETH, led by Andrea Rubbia and many others like the colleagues of Granada University led by Antonio Bueno. In particular we want to honor the merit of all the technical staff of INFN who validly contributed to the T600 realization with their highly professional skill. We cannot even forget the contribution and joint effort of many industrial realities that played a major role in the realization of the detector: Air Liquid and Stirling for cryogenics, Breme Tecnica, Galli-Morelli and CINEL for the inner detector, and CAEN for the data acquisition electronics. The Polish groups acknowledge the support of the Ministry of Science and Higher Education in Poland for four SPB grants, the INFN FAI program and the EU Commission. Finally we thank CERN, in particular the CNGS staff, for the successful operation of the neutrino beam facility, which represents an important part of the ICARUS physics programme.